\documentstyle[epsf]{preprint}
\journalid{*}{*}{1998}
\articleid{1}{6}

\title{FOSSIL IMPRINTS OF THE FIRST GENERATION SUPERNOVA EJECTA IN
EXTREMELY METAL-DEFICIENT STARS} 
{FOSSIL IMPRINTS OF THE FIRST GENERATION SUPERNOVA EJECTA} 

\author{Toshikazu Shigeyama\\
{\em Department of Astronomy, School
of Science, University of Tokyo, Bunkyo-ku, Tokyo, 113-0033 Japan}\\
{\em Research Center for the Early Universe, School of Science, University
of Tokyo,  Bunkyo-ku, Tokyo, 113-0033 Japan;} \\
{\em shigeyama@astron.s.u-tokyo.ac.jp} \\
 Takuji Tsujimoto \\
{\em National Astronomical Observatory, Mitaka, Tokyo, 181-8588 Japan;} \\
{\em tsuji@misty.mtk.nao.ac.jp}}
{T. Shigeyama \& T. Tsujimoto}

\received{8 July 1998}
\accepted{14 September 1998}

\abstract{
Using results of nucleosynthesis calculations for theoretical
 core-collapse supernova models with various progenitor's masses, it
 is shown that abundance patterns of C, Mg, Si, Ca, and H seen in
 extremely metal-deficient stars with [Fe/H] \ltsim --2.5 follow those
 seen in the individual first generation supernova remnants
 (SNRs). This suggests that most of the stars with [Fe/H] \ltsim --2.5
 were made from individual supernova (SN) events. To obtain the ratio
 of heavy elements to hydrogen, a formula is derived to estimate the
 mass of hydrogen swept up by a SNR when it occurs in the interstellar
 matter with the primordial abundances. We use [Mg/H] to indicate the
 metallicities instead of [Fe/H]. The metallicities [Mg/H] predicted
 from these SNRs range from $\sim -4$ to $\sim -1.5$ and the mass of
 Mg in a SN is well correlated with its progenitor's mass. Thus the
 observed [Mg/H] in an extremely metal deficient star has a
 correspondence to the progenitor's mass. A larger [Mg/H] corresponds
 to a larger progenitor's mass. Therefore, so called `age-metallicity
 relation' does not hold for stars with [Fe/H] \ltsim --2.5. In
 contrast, the [Mg/Fe] ratios in the theoretical SNRs have a different
 trend from those in extremely metal-deficient stars.  It is also
 shown that the observed trend of [Mg/Fe] can predict the Fe yield of
 each SN given the correspondence of [Mg/H] to the progenitor's
 mass. The Fe yields thus obtained are consistent with those derived
 from SN light curve analyses. This indicates that there is still a
 problem in modelling a core-collapse supernova at its beginning of
 explosion or mass cut. The abundance determination of O in extremely
 metal-deficient stars, that have not been done from observational
 analyses, are strongly desired to test the hypothesis that the
 elements in an extremely metal-deficient star come from a single SN
 event and to obtain reliable yields for SNe.
}


\def\etal{{et al. }}
\def\Msun{M_{\odot}}
\def\cm3{{\rm ~cm}^{-3}}
\def\ltsima{$\; \buildrel < \over \sim \;$}
\def\ltsim{\lower.5ex\hbox{\ltsima}}
\def\gtsima{$\; \buildrel > \over \sim \;$}

\begin{document}
\section{INTRODUCTION}
Recent observations and the analyses imply that the abundance pattern
of an extremely metal-deficient star with
[Fe/H]$\left(\equiv\log\left({\rm Fe/H}\right)-\log\left({\rm
Fe/H}\right)_\odot\right)$ \ltsim $-2.5$ may retain information of a
preceding single supernova (SN) event or at most a few SNe
(\cite{mcwilliam95xx}; \cite{ryan96}). A theoretical attempt from this
point of view has been already made (\cite{audouze95}) to explain the
observed abundance patterns. They argued that a combination of yields
from two SNe with different progenitor's masses at the main sequence
($M_{\rm ms}$) is consistent with the abundance patterns of stars with
the lowest metallicity ([Fe/H]$\sim -4$).

Assuming that the formation of extremely metal-deficient stars is
triggered by a single supernova remnant (SNR) and that the formed
stars retain the abundance pattern of this SN, one could predict
abundance patterns (including hydrogen) of these stars from
theoretical SN models available at present (e.g., \cite{woosley95};
\cite{tsujimoto95}; \cite{nomoto97}). In this {\it letter}, we will
compare the abundance pattern thus obtained with observations to test
whether the chemical enrichment by individual SNRs could explain the
observed abundances on the surfaces of extremely metal-deficient
stars. We will use models in \cite{woosley95x} (referred to WW95) and
those in \cite{tsujimoto95x} and \cite{nomoto97x} (T95).

\begin{figure}[ht]
\begin{center}
\leavevmode
\epsfxsize=\columnwidth\epsfbox{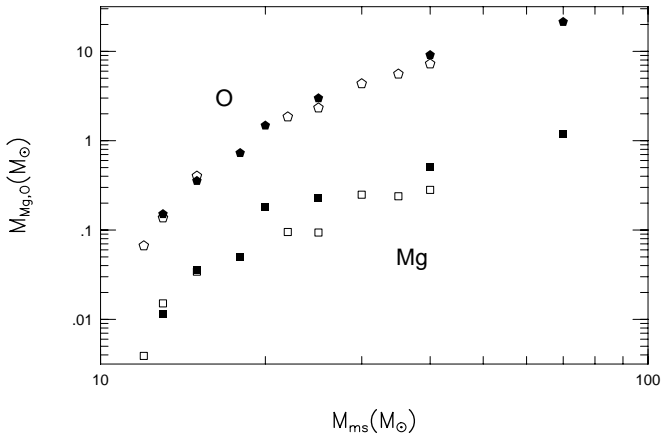}\hfil
\end{center}
\caption{The masses of Mg (circles)
and O (pentagons) ejected from SN models are plotted as a 
function of the progenitor's mass (\protect{\cite{woosley95} }(open);
\protect{\cite{tsujimoto95}} (filled)). 
\label{mgm}}
\end{figure}

The metallicity of a star has been usually indicated by [Fe/H]. On the
other hand, the yields of Fe from SN models to date have not converged
(\cite{woosley95}; \cite{tsujimoto95}) because of uncertainties in the
explosion mechanism and fall back dynamics or mass cut between the
forming neutron star (or black hole) and the ejected envelope (e.g.,
\cite{thielemann90}). Thus the yields of lighter $\alpha$-elements are
more reliably calculated. We will use [Mg/H] instead of [Fe/H] to
specify the metallicity because of the following reasons:
\begin{enumerate}
\item Mg is less affected by the mass cut in SN models than Fe.
\item Mg is not synthesized or broken by the SN shock.
\item The mass of ejected Mg increases with increasing
progenitor's mass.
\item The abundance of Mg is observationally derived for many
stars with [Fe/H] \ltsim $-2.5$.
\end{enumerate}
There is a disadvantage in this element, that is, two SN models (WW95
and T95) give somewhat different masses of Mg as a function of
progenitor's mass (squares in Fig. \ref{mgm}). The best element to be
used in this respect would be O (pentagons in Fig. \ref{mgm}) if O
abundances were available for many stars with [Fe/H] \ltsim $-2.5$
(see \cite{mcwilliam95xx}; \cite{ryan96x}).

\begin{figure}[ht]
\begin{center}
\leavevmode
\epsfxsize=0.9\columnwidth\epsfbox{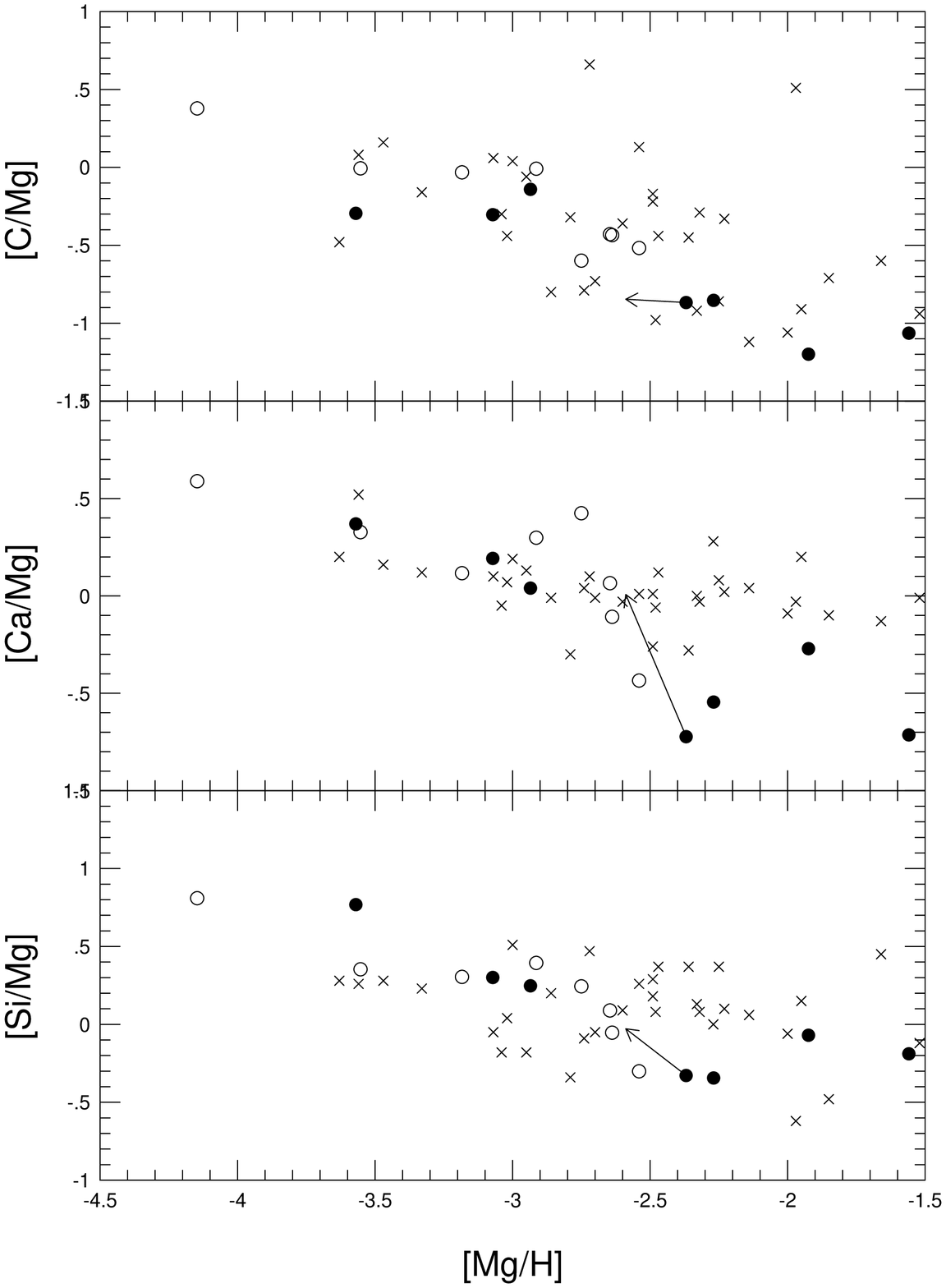}\hfil
\end{center}
\caption{The top panel: the crosses are the observed [C/Mg]
for stars plotted against the [Mg/H]
(\protect{\cite{mcwilliam95xx}}). The open and filled circles show the
same quantities in the first generation SNRs calculated from
theoretical SN models (\protect{\cite{woosley95} }(open circles);
\protect{\cite{tsujimoto95}} (filled circles)). The arrow indicates
the change in the abundance pattern of the model to reproduce the SN
1987A observations (\protect{\cite{thielemann90x}} and references
therein). The middle panel: Same as the top panel but for [Ca/Mg]. The
bottom panel: Same as the top panel but for [Si/Mg]. See the text for
arrows.
\label{cmg}}
\end{figure}

There is a clear trend in the observed [C/Mg]$-$[Mg/H] plot (crosses
in Fig. \ref{cmg}) for [Mg/H]$<-2$ that cannot be reconciled with the
conventional one-zone models of the Galactic chemical evolution that
assume a complete mixing of elements inside the
Galaxy.\footnote[1]{However, \cite{audouze95x} have deduced that the
chemical inhomogeneity is maximal in the range $-4\ltsim$[Fe/H]$\ltsim
- 2.5$ ($-3.5\ltsim$[Mg/H]$\ltsim - 2$).}  The difference of the
observed [C/Mg] in the range of $-3.5$\ltsim[Mg/H]\ltsim$ -2$ gives a
mean gradient $ \Delta {\rm [C/Mg]}/ \Delta {\rm [Mg/H]}\sim -
0.7$. On the other hand, the one-zone Galactic chemical evolution
model using T95 would predict the evolutionary change in [C/Mg]
starting with the value of $\sim -1.2$ given by the most massive star
$M_{\rm ms}\sim 50\Msun$ and converging to the average value of $\sim
- 0.9$ integrated for stars in the mass range of $10<M_{\rm
ms}/\Msun<50$ over the Salpeter initial mass function (IMF) with a
slope of $x=-1.35$. It results in the corresponding gradient of $
\Delta {\rm [C/Mg]}/ \Delta {\rm [Mg/H]}\sim +0.2$ that has the
opposite sign and a smaller absolute value than in observations. This
argument also holds for WW95. If we introduce a metallicity-dependent
IMF, the observed gradient requires a drastic change of the slope of
the IMF from $x\sim -10$ at [Fe/H]$\sim - 4$ to $x\sim+0.5$ at
[Fe/H]$\sim -2.5$. The index of $x\sim-10 $ is almost equivalent to
assuming that most of the stars with [Fe/H]$\sim - 4$ are formed from
SNRs with $M_{\rm ms}=10\,\Msun$. It will be shown in the following
section that this is indeed the case if we assume that most of the
stars with [Fe/H]$\ltsim -2.5$ are formed from individual SN events.

 Here we will also predict the Fe yields as a function of $M_{\rm ms}$
using the observed [Mg/Fe]--[Mg/H] trend combined with the
[Mg/H]$-M_{\rm ms}$ relation in theoretical SN models. Then these Fe
yields are compared with those derived from SN light curve analyses.
To calculate the mass of hydrogen swept up by a SNR, that is important
to obtain [Mg/H], the analytical expression for this mass in
\cite{cioffi88} is modified to be applicable to SNRs that occur in a
metal-free interstellar matter (ISM).

\section{LATE STAGE EVOLUTION OF THE FIRST GENERATION SUPERNOVA REMNANTS}

%\subsection{Scenario toward the formation of stars of the next generation}
In a spherically symmetric SNR (e.g., \cite{cioffi88x}), after the
radiative cooling time scale becomes shorter than the dynamical time
scale at the shock front, the shock front propagates considerably
slower than before and makes a dense shell immediately behind it. Most
of the mass inside the shock front resides in this shell. The ejecta
that contain heavy elements occupy only a small central portion of the
SNR even during this pressure driven snowplow (PDS) phase. In a
realistic SNR, the situation is different. First, the contact surface
between the ejecta and the ISM is subject to the Rayleigh-Taylor
instability during the preceding Sedov-Taylor phase (\cite{taylor50};
\cite{sedov59}). A large part of the ejecta penetrates into the
ISM. Thus it is expected that the ejecta already approach the dense
shell at the beginning of the PDS phase and merge with it during this
phase. Secondly, an isothermal shock front is considered to be
dynamically unstable (\cite{vishniac83}). Thus the dense shell behind
the shock front is fragmented into a number of cloud cores that retain
the abundance pattern of the SN. These cloud cores embedded in a high
ambient pressure become seeds of stars of the next generation
(\cite{nakano98}). Some of the stars thus formed are currently
observed as extremely metal-deficient stars. Accordingly, the average
abundance pattern inside the SNR is assumed to represent that in all
these stars.

%\subsection{Life time of SNR}
\cite{cioffi88xx} obtained the time $t_{\rm merge}$ when a SNR
loses its identity. This time is a function of the ratio of the shock
velocity to the sound speed $C_{\rm S}$ (or the velocity dispersion)
of the ISM at the beginning of PDS phase ($t=t_{\rm PDS}$) and $t_{\rm
PDS}$ itself. For a SNR in a homogeneous ISM with the density
of $n_1$, $t_{\rm PDS}$ is determined by equation
\begin{equation}\label{pds}
{1 \over n_1}{\partial e \over \partial t}=-\alpha \times \Lambda (T),
\end{equation}
where $e$ and $\Lambda(T)$ are the internal energy and the cooling
function at the temperature $T$, respectively. All the variables are
evaluated at the shock front using Sedov-Taylor solutions for the
point explosion with the energy $E_0$ and the number density
$n_1$. The constant $\alpha$ should be equal to 1.85 to reproduce
$t_{\rm PDS}$ for the numerical simulation in \cite{cioffi88xx} when
$\Lambda(T)$ for the solar abundances is used. Then $\Lambda_{\rm
primordial}(T)$ is constructed for the gas composed of only hydrogen
and helium with their mass ratio $X_{\rm H}: X_{\rm He}=0.75:0.25$
under the collisional ionization equilibrium. For
$n_1$\gtsima$10^{-2}$ cm$^{-3}$, equation (\ref{pds}) gives $T$
greater than 10$^5$ K. Thus the ionization of hydrogen does not affect
the dynamics.

%\subsection{Mass of hydrogen swept up by a supernova}
The mass of hydrogen $M_{\rm sw}$ thus obtained with
$\Lambda(T)=\Lambda_{\rm primordial}(T)$ in equation (\ref{pds}) is
approximated by the formula
\begin{eqnarray}\label{swept}
&M_{\rm sw}=5.1\times 10^4 \Msun &\nonumber \\
&\times {\left({E_0 \over 10^{51} {\rm
erg}}\right)}^{0.97} n_1^{-0.062}&\left({C_{\rm S}\over 10\,{\rm
km\,s}^{-1}}\right)^{-9/7} .
\end{eqnarray}
The mass $M_{\rm sw}$ is insensitive to $n_1$. The sound speed $C_{\rm
S}$ was assumed to be 10 km s$^{-1}$ (or $T\sim 10^4$ K).  This mass
depends on $E_0$ and $n_1$ in a different way from that of
\cite{cioffi88xx} due to the different cooling function used here. The
cooling function with the primordial abundances is approximately
proportional to $T^{-2}$ instead of $T^{-1/2}$ in \cite{cioffi88xx}.

\section{ABUNDANCE PATTERNS}
\label{alpha}
Here we assume that all the stars with [Mg/H]$<-2$ are made from
individual SNRs.  Abundance patterns in individual SNRs will be
calculated based on yields from theoretical SN models and compared
with those on the surface of metal-deficient stars obtained by
\cite{mcwilliam95x}. For SN models in metal-free environment, we use
models Z12A, Z13A, Z15A, Z22A, Z30B, Z35C, Z40C in WW95 to retain the
monotonicity of the mass of Mg as a function of progenitor's mass (see
filled squares in Fig. \ref{mgm}). Another group of SN models are
taken from T95 in which the initial metallicity is solar. Parameters
of both models are listed in Table \ref{abund}.

\begin{table*}
\begin{center}
\caption{Input parameters in SN models and the calculated [Mg/H] ratios in the corresponding SNRs (see text for details).}
\begin{tabular}{cccc||ccc} \hline \hline
\multicolumn{4}{c||}{WW95} & \multicolumn{3}{c}{T95} \\
Name &$M_{\rm ms}$ ($\Msun$)  &
$E_0$ ($10^{51}$ erg) &[Mg/H] & $M_{\rm ms}$ ($\Msun$)&$E_0$ ($10^{51}$ erg) & [Mg/H]\\
\hline 
Z12A & 12  & 1.28 & -4.1 &13 & 1 & -3.6 \\
Z13A & 13  & 1.29 & -3.6 &15 & 1 & -3.1 \\
Z15A & 15  & 1.27 & -3.2 &18 & 1 & -2.9 \\
Z22A & 22  & 1.26 & -2.8 &20 & 1 & -2.4 \\
Z25B & 25  & 1.83 & -2.9 &25 & 1 & -2.3 \\
Z30B & 30  & 2.06 & -2.5 &40 & 1 & -1.9 \\
Z35C & 35  & 2.49 & -2.6 &70 & 1 & -1.6 \\
Z40C & 40  & 3.01 & -2.6 &- & - & - \\
\hline
\end{tabular}
\label{abund}
\end{center}
\end{table*}

We divide elements into two categories. One category is composed of
the elements not (or at least less) affected by the mass cut in SN
models. The examples are C, Mg, Si, and Ca. The other is composed of
those affected by the mass cut and includes Cr, Mn, Fe, Co, Ni, for
example. This division is necessary because the mass cut in SN models
to date is totally artificial and uncertain.

{\sl Elements not influenced by mass cut}--- The yields of elements
not influenced by the mass cut are expected to be more reliably
calculated in SN models than those influenced by the mass
cut. Therefore observed abundance ratios of [C/Mg], [Ca/Mg], [Si/Mg]
with respect to [Mg/H] (Fig. \ref{cmg}) can tell whether extremely
metal-deficient stars were formed from individual SNe. These abundance
ratios derived from SN models (open circles:WW95; filled circles: T95)
are also plotted in the same figure, in which [Mg/H] is calculated
from the ratio of the mass of Mg in a SNR to that of hydrogen swept up
by the same SNR (see equation (\ref{swept})). The relations between
[Mg/H] and $M_{\rm ms}$ are shown in Table 1 for both SN models. The
values of [Mg/H] for $M_{\rm ms}\sim 10\Msun$ (the lower mass limit of
the core-collapse SN progenitor) from both models are not so far from
the observed lowest value of [Mg/H]$\sim -3.7$ and the metallicity
[Mg/H] increases with increasing progenitor mass $M_{\rm ms}$. This
infers that there are more stars at lower metallicities according to
an IMF that decreases toward high masses if stars with lower masses,
say $10\,\Msun$, explode in the regions that have not been polluted by
other SNRs before. This condition is represented as $n_\star V_{\rm
SNR}\int_{10 \Msun}^{50\,\Msun}\phi(m)dm\, <\,1\, {\rm star}$. Here
$n_\star$, $\phi$, and $V_{\rm SNR}$ denote the number density of the
first generation stars, the IMF normalized to unity between the lower
and upper mass limits, and the maximum volume occupied by a single
SNR, respectively. This condition results in a star formation
efficiency for the first generation stars smaller than 2\%, if the
Salpeter IMF is used.

A trend seen in the observed [C/Mg] (crosses in the top panel of
Fig. \ref{cmg}) is well reproduced by both models. This means that
stars with [Mg/H]$\sim -3.5$ were made from individual low mass
($M_{\rm ms}\sim 10\,\Msun$) SNe and that stars with [Mg/H]$\sim -2$
were from individual high mass SNe ($M_{\rm ms}\sim 50\,\Msun$). For
[Si/Mg], both models are also consistent with observations, though the
observed [Si/Mg] has no clear trend. The [Ca/Mg] ratios derived from
T95 with $M_{\rm ms}\geq 20 \Msun$ deviate from observations and gives
lower values in the region [Mg/H]$>-2.5$. This indicates that the
yields of Ca of T95 with $M_{\rm ms}\geq 20 \Msun$ are too
small. These abundance patterns together with the predicted [Mg/H]
ratios in Table 1 suggest that stars with $-3.5\ltsim{\rm
[Mg/H]}\ltsim -2$ are made from individual SN events.

\begin{figure}[ht]
\begin{center}
\leavevmode
\epsfxsize=0.9\columnwidth\epsfbox{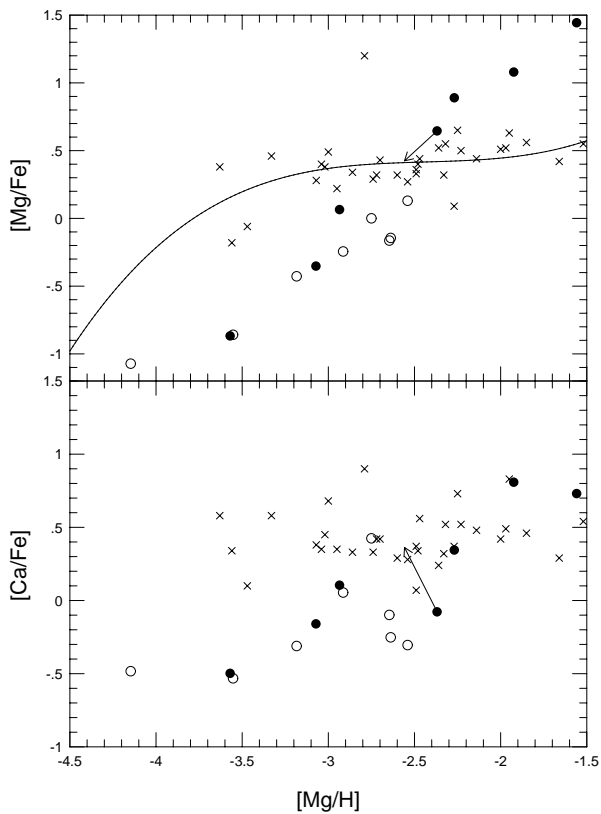}\hfil
\end{center}
\caption{The upper panel: the crosses are the observed
[Mg/Fe] for stars plotted against the [Mg/H]
(\protect{\cite{mcwilliam95xx}}). The solid curve shows the
$\chi$-square fit of the observed points with the errors given in \protect{\cite{mcwilliam95x}} to a cubic polynomial formula 
. The open and filled circles show the same quantities in the first generation SNRs
calculated from theoretical SN models
(\protect{\cite{woosley95} }(open circles);
\protect{\cite{tsujimoto95}} (filled circles)). The lower panel: Same
as the top panel but for [Ca/Fe]. See the text for arrows.
\label{femg}}
\end{figure}

{\sl Elements influenced by mass cut}--- The upper panel of
Fig. \ref{femg} shows that the [Mg/Fe] ratios derived from both models
do not follow the observations. This reconfirms that none of the two
models correctly describes the dynamics and/or nucleosynthesis near
the surface of the Fe core in progenitor stars during explosion. In
both models, less massive stars have too large Fe yields.  More
massive stars in T95 have too small Fe yields but the opposite is the
case for WW95. The lower panel of Fig. \ref{femg} shows that the
[Ca/Fe] ratios from both models are smaller than the
observations. This can be ascribed to too large Fe yields from less
massive stars of theoretical models again. An apparent fit of the
observational data to the model for [Mg/H]$>-2.5$ is a consequence of
two errors in yields of both Ca (see section \ref{alpha}) and Fe for
more massive stars (see the next section).

Abundances of some heavy elements in the ejecta of the 20 $\Msun$
supernova were deduced from the SN 1987A observations
(\cite{thielemann90x} and references therein). As the mass of Mg
ejected from SN 1987A was not estimated from observations, we derive
the mass of Mg by reducing the value of T95 to match the observed
[Mg/Fe] ratio in extremely metal-deficient stars.  The corresponding
change of [Mg/Fe] is shown by the arrow in the upper panel of
Fig. \ref{femg}. If we use the mass of Mg thus derived and the
observed masses of C, Si, Fe, and Ca (the upper limit), the abundance
ratios of the model with $M_{\rm ms}=20\,\Msun$ would move following
the arrows in Fig. \ref{cmg} and in the lower panel of
Fig. \ref{femg}. Thus the abundance pattern of the best observed SN
coincides with that in extremely metal-deficient stars with
[Mg/H]$\sim-2.6$ and the mass of Mg in the 20 $\Msun$ model of T95
might be overestimated.

\section{IRON YIELDS DEDUCED FROM EXTREMELY METAL DEFICIENT STAR OBSERVATIONS}

\begin{figure}[ht]
\begin{center}
\leavevmode
\epsfxsize=\columnwidth\epsfbox{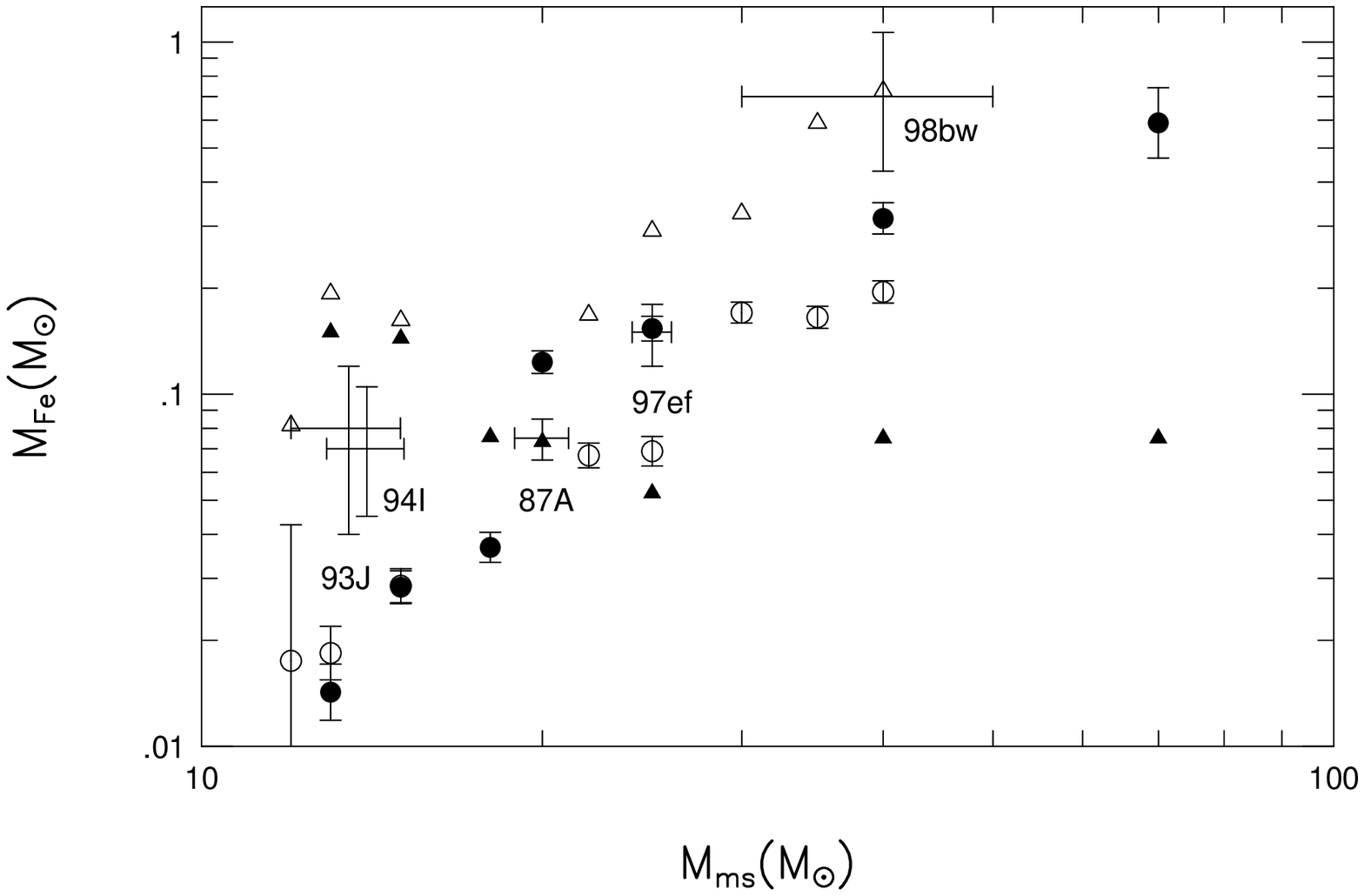}\hfil
\end{center}
\caption{The masses of Fe in SN models
(\protect{\cite{woosley95} }(open); \protect{\cite{tsujimoto95}}
(filled)) are shown by triangles. Circles show the masses of Fe in SN
models with 1 $\sigma$ error bars of the $\chi^2$ fit in the upper
panel of Fig. \protect{\ref{femg}}. These masses are determined by the
[Mg/Fe]$-$[Mg/H] relations indicated with the solid curve in the upper
panel of Fig. \protect{\ref{femg}} and the [Mg/H]$-M_{\rm ms}$
relations. Points with error bars are the masses of $^{56}$Ni, that
eventually decays to $^{56}$Fe derived from SN light curve analyses
(SN 1987A: \protect{\cite{shigeyama90}}; SN 1993J:
\protect{\cite{shigeyama94}}; SN 1994I: \protect{\cite{iwamoto94}}: SN
1997ef: \protect{\cite{iwamoto98a}}; SN 1998bw:
\protect{\cite{iwamoto98b}}).
\label{fem}}
\end{figure}

We will reverse the argument in the preceding sections. Suppose that
the trends in the observed abundance ratios, say [Mg/Fe], tell yields
of SNe as a function of [Mg/H].  The $\chi^2$ fit of the observational
data to a cubic polynomial formula gives the solid curve in the upper
panel of Fig. \ref{femg}. This curve and [Mg/H] of each model predict
the mass of Fe ($M_{\rm Fe}$) from a SN as a function of $M_{\rm
ms}$. Filled circles in Fig. \ref{fem} show the predicted Fe masses
using T95. These Fe masses are completely different from those given
in the original models (triangles) but hardly change the average Fe
yield integrated over the Salpeter IMF. Open circles are those from
WW95 and give a similar amount of Fe for each progenitor's
mass. Points with error bars also shown in the same figure are the Fe
masses obtained from SN light curve analyses. Good fits of the Fe
masses from the light curve analyses to those inferred from the
abundance patterns demonstrate that this procedure to obtain the Fe
yields works, if the Fe yields are not affected by the initial
metallicities.  A discrepancy of Fe masses at $M_{\rm ms}\sim
10\,\Msun$ might be due to the small number of fairly scattered
observed points of [Mg/Fe] at [Mg/H]$\sim-3.5$ in the upper panel of
Fig. \ref{femg}.  The Fe yields thus obtained predict that the lowest
[Fe/H] should be $\sim -3.7$ for both models.

\section{CONCLUSIONS} 

We have shown that the abundance patterns of C, Mg, Si, Ca, and H
theoretically predicted in the first generation SNRs are in good
agreement with those observed in extremely metal-deficient stars with
$-4<$[Fe/H]$<-2.5$.  All these elements are thought to be less
affected by the mass cut in SN modelling than heavier elements like
Fe. This agreement implies that each extremely metal-deficient star
was formed from a single SN event. In contrast, both theoretical SN models 
predict different trends in the
[Mg/Fe]$-$[Mg/H] plot from the observed one. Conversely, the mass of
Fe ejected by each SN as a function of $M_{\rm ms}$ can be derived
from the observed [Mg/Fe]$-$[Mg/H] trend combined with the
[Mg/H]$-M_{\rm ms}$ relation in theoretical SN models. This $M_{\rm
Fe}$ as a function of $M_{\rm ms}$ is consistent with that derived
from SN light curve analyses. Following the same procedure presented
in this {\it letter} to obtain the Fe yields, we can derive the yields
for other elements such as Ti, Cr, Mn, Co, Ni and r-process elements
(\cite{tsujimoto98}), all of which are very uncertain in SN models. If
O abundances are deduced in extremely metal-deficient stars, we will
be able to obtain more reliable yields for SNe. In the forthcoming
paper we will also discuss the connection between the chemical
enrichment by individual SNe and the subsequent chemical evolution.

%\acknowledgements 
~

We are grateful to the anonymous referee for making useful comments. This work has been partially supported by COE research (07CE2002) of
the Ministry of Education, Science, Culture, and Sports in Japan.

\end{document}